\documentclass[twocolumn,superscriptaddress]{revtex4-1}

\usepackage{graphicx}
\usepackage{amsmath}
\usepackage{amsfonts}
\usepackage{amssymb}
\usepackage[normalem]{ulem}
\usepackage[svgnames]{xcolor}
\usepackage{bbm}
\usepackage{bm}

\newcommand{\Ag}{A_g}
\newcommand{\Au}{A_u}
\newcommand{\BIg}{B_{1g}}
\newcommand{\BIIg}{B_{2g}}
\newcommand{\BIIIg}{B_{3g}}
\newcommand{\BIu}{B_{1u}}
\newcommand{\BIIu}{B_{2u}}

\begin{document}
\title{Valence force model and nanomechanics of single-layer phosphorene}

\author{Daniel Midtvedt}
\email{midtvedt@chalmers.se}
\affiliation{Department of Physics, Chalmers University of Technology, Gothenburg, Sweden}
\author{Alexander Croy}
\email{croy@pks.mpg.de}
\affiliation{Max-Planck-Institute for the Physics of Complex Systems, Dresden, Germany}
\begin{abstract}
In order to understand the relation of strain and material properties, both a microscopic model connecting a given strain to the displacement of atoms, and a macroscopic model relating applied stress to induced strain, are required. Starting from a valence-force model for black phosphorous (phosphorene) [Kaneta \textit{et al., Solid State Communications}, 1982, {\bf 44}, 613] we use recent experimental and computational results to obtain an improved set of valence-force parameters. From the model we calculate the phonon dispersion and the elastic properties of single-layer phosphorene. Finally, we use these results to derive a complete continuum model, including the bending rigidities, valid for long-wavelength deformations of phosphorene. This continuum model is then used to study the properties of pressurized suspended phosphorene sheets.
\end{abstract}
\maketitle

\section{Introduction}
Phosphorene (or black phosphorus) has attracted a lot of
interest in recent times\cite{chja14,liwa+15}. Its unusual puckered structure leads to anisotropic elastic and (opto)electronic 
properties\cite{xiwa+14,feya14a,qiko+14,wepe14,waku+15,wajo+15}, which is interesting for applications and strain-engineering purposes\cite{liyu+14,bugr+14,Roldan2015}.
Accordingly, there are many computational and experimental studies of the
properties of phosphorene. As its structure is the key for understanding the anisotropy of
the material, several \emph{ab initio} calculations of the phonon and elastic properties
of phosphorene have been performed. However, for some questions the use of \emph{ab initio} methods is not feasible.
In such situations, so-called valence-force models (VFMs) provide a viable alternative due to their relative
computational simplicity. When accurately parametrized, they can be used to extract the phonon dispersion and elastic properties.
Moreover, VFMs can be used to connect the microscopic structure and macroscopic quantities such as strain\cite{mile+16}.
In particular, atomic displacements can be related to the macroscopic strain applied to the material,
which is a requirement for strain-engineering investigations\cite{Roldan2015}.

In this work we present a VFM for phosphorene. Our model is based on the VFM originally given in Ref.\ \cite{kaka+82}.
By using several experimental\cite{sush85,suue+81,lash79,akko+97,cavi+14} and recent computational results\cite{elkh+15,qiko+14,wepe14,sali+14,waku+15} and constraints, we re-optimize the VFM parameters.
We obtain the elastic constants and the bending rigidities from the VFM, which allows us to construct a complete
continuum mechanics model. We apply this model to a pressurized and suspended phosphorene drum and derive deflection and
strain distributions.

\section{Model}
\subsection{Structure}
\begin{figure}[b!]
  \centering
  \includegraphics[width=0.4\textwidth]
             {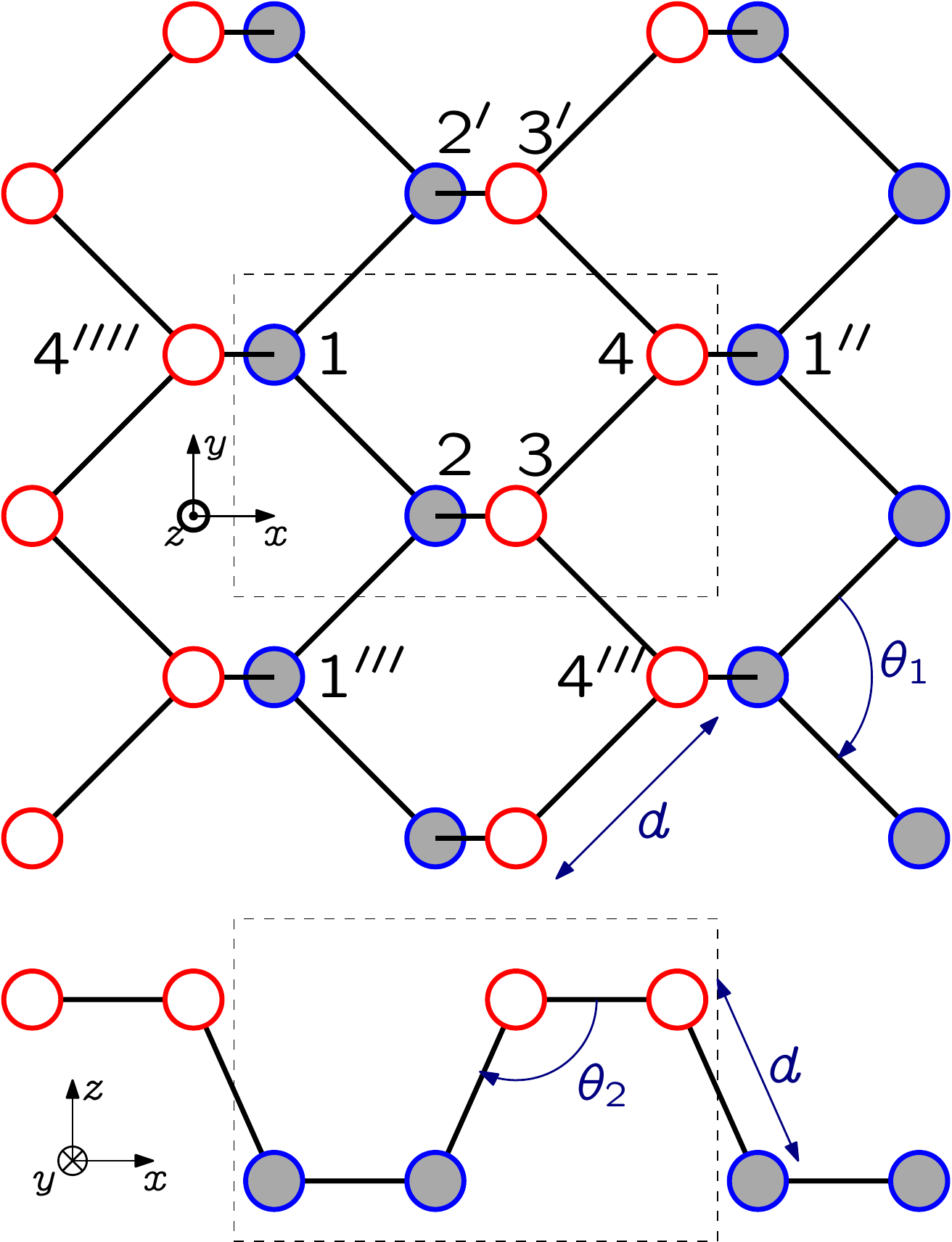}           
             \caption{Lattice structure of phosphorene. The unit cell is indicated by a dashed rectangle and contains the atoms
                $1,2,3,4$.}
\label{fig:lattice}
\end{figure}
On an atomic level phosphorene consists of an orthorhombic lattice with lattice vectors $\vec{a}_1$ and $\vec{a}_2$ and four basis atoms arranged in a puckered structure as depicted in Fig.\ \ref{fig:lattice}. We denote the atomic positions by $\vec{r}_i$ with subindex $i$ running from $1$ to $N$ where $N$ is the number of atoms in the phosphorene sheet. Displacements of these positions are denoted by $\delta \vec{r}_{i}$. The inter-atomic bond vectors are then given by $\vec{r}_{ij}=\vec{r}_{j}-\vec{r}_i$ and the angle defined by three atoms $j,i,k$ with atom $i$ as the vertex is $\theta_{jik}$. The equilibrium structure is characterized by the inter-atomic spacing $d\approx 2.22 {\rm \AA}$ and intra- and inter-pucker angles $\theta_1\approx 96.5^{\circ}$ and $\theta_2 \approx 101.9^{\circ}$ \cite{kaka+86,Note1}.
The equilibrium bond vectors thus become $\vec{r}_{12} = d\left( \cos(\theta_1/2),  -\sin(\theta_1/2), 0 \right)$,
$\vec{r}_{23} = d\left(-\cos(\theta_2)/\cos(\theta_1/2), 0, \sqrt{1- \left(\cos(\theta_2)/\cos(\theta_1/2)\right)^2} \right)$
and
$\vec{r}_{34} = d \left(  \cos(\theta_1/2),  \sin(\theta_1/2), 0 \right)$. The lattice vectors are 
$\vec{a}_1 = 2 d (\cos(\theta_1/2) -\cos(\theta_2)/\cos(\theta_1/2), 0, 0)$ and $\vec{a}_2 = 2 d (0, \sin(\theta_1/2), 0)$.

\subsection{Valence force model}
To model distortions of the phosphorene lattice we adopt the valence force model originally proposed in Ref.\ \cite{kaka+82}. Accordingly, the deformation energy is given by
\begin{align}
    \mathcal{E}_{\rm def} ={}& \frac{1}{2}\sum_i\left(
        \frac{1}{2}\sum_{j\in_{\rm ip} i}d^2K_r \delta r_{ij}^2 
        + \frac{1}{2}\sum_{j\in_{\rm cp} i}d^2K'_r \delta r_{ij}^2 \right. \nonumber \\ 
    &    +\sum_{k < j \in_{\rm ip} i} d^2 K_{\theta} \delta \theta_{jik}^2  
       +\sum_{j\in_{\rm ip} i}\sum_{k\in_{\rm cp} i} d^2 K'_{\theta}\delta \theta_{jik}^2 \nonumber\\
    &    +\sum_{k < j \in_{\rm ip} i} d^2 K_{rr'} \delta r_{ij} \delta r_{ik}  \nonumber \\
    &   +\sum_{j\in_{\rm ip} i}\sum_{k \in_{\rm cp} i} d^2 K'_{rr'}\delta r_{ij} \delta r_{ik} \nonumber\\
    &    +\sum_{k < j \in_{\rm ip} i} d^2 K_{r\theta}(\delta r_{ij} + \delta r_{ik} )\delta \theta_{jik}   \nonumber \\
 &\left.+\sum_{j\in_{\rm ip} i}\sum_{k \in_{\rm cp} i} d^2 (K'_{r\theta} \delta r_{ij} + K''_{r\theta} \delta r_{ik} )\delta \theta_{jik}
\right). \label{eq:vfm_en}
\end{align}
In this expression, $\delta r_{ij}=|\vec{r}'_{ij} - \vec{r}_{ij}|/d \approx (\vec{r}'_{ij} - \vec{r}_{ij})\cdot\vec{r}_{ij}/d^2$ 
is the relative change in bond-length between atoms $i$ and $j$ and 
$\delta \theta_{ijk} \approx - (\cos(\theta_{ijk}' )- \cos(\theta_{ijk}))/\sin(\theta_{ijk})$ 
with $\cos(\theta_{ijk}')\approx (\vec{r}'_{ij}\cdot \vec{r}_{jk}')(1-\delta r_{ij}/d-\delta r_{jk}/d)/d^2$ is the change in angle between atoms $i,j,k$ with atom $j$ as apex. The sum over $i$ runs over all atoms in the sheet. The sums over $j\in_{\rm ip} i$ run over nearest neighbors to atom $i$ within the same pucker. This leaves two terms for each atom. The terms which contain a sum over $k < j\in_{\rm ip} i$ are constructed out of both neighbors of atom $i$ that belong to the same pucker. Thus, this sum consists of a single term. Finally, the sum over $k\in_{\rm cp} i$ just contains the single atom $k$ neighboring $i$ that belongs to a different pucker. In total there are nine force-field parameters, namely, $K_r$, $K'_r$, $K_{\theta}$, $K'_{\theta}$, 
$K_{rr'}$, $K'_{rr'}$, $K_{r\theta}$, $K'_{r\theta}$ and $K''_{r\theta}$. 
Those determine the energy cost for bond stretching, angle bending, bond-bond and 
bond-angle correlations, respectively.

\section{Results and discussion}
\subsection{Phonon frequencies}
From the energy \eqref{eq:vfm_en} we obtain the dynamical matrix, which allows us to calculate the
phonon dispersion of phosphorene. For each momentum vector $\vec{k}$ there are $4\times 3$ phonon modes. Three of them are acoustic modes
with a vanishing frequency at the $\Gamma$-point $\vec{k}=\vec{0}$. From group theory\cite{rial+15} it follows, that six of the remaining
modes are Raman active ($\Ag^1$, $\Ag^2$, $\BIg$, $\BIIg$, $\BIIIg^1$ and $\BIIIg^2$), two are infrared active ($\BIu$ and $\BIIu$)
and one mode is silent ($\Au$) (we use the notation of Ref.\ \cite{feya14a}). 

As we will show, one can use the frequencies of the six Raman modes and one of the infrared modes at the $\Gamma$-point to fix seven of the
force-field parameters essentially without fitting. The remaining two parameters do not influence the frequencies of the optical modes,
but can be used to adjust the elastic properties. To express the frequencies at the $\Gamma$-point in terms of the force-field
parameters, we project the dynamical matrix onto the group-theoretical eigenvectors. The resulting matrix is block-diagonal and the respective frequencies can easily be obtained.

In the following we describe the general procedure to get seven force-field parameters. This procedure can be implemented numerically
or by using the analytic expressions given in the supplement. Once a parameter value is found, it is used in the subsequent calculations.
As a first observation, we find that the frequencies of $\BIg$ and $\BIu$ are both determined by $K'_{\theta}$ alone. The ratio $\omega^2_{\BIu}/\omega^2_{\BIg} \approx 1.62$ is fixed by the equilibrium geometry. This means we can only use one of the modes to find the value of $K'_{\theta}$. Having $K'_{\theta}$ we use the frequencies of $\BIIIg^1$ and $\BIIIg^2$ to obtain the values of
$K'_r$ and $K''_{r\theta}$. We proceed by choosing (arbitrary) values for $K_{\theta}$ and $K_{r\theta}$, for example $K_{\theta}=K'_{\theta}$ and $K_{r\theta}=K''_{r\theta}$. Then we obtain
$K_r + K_{rr}$, $K'_{rr'}$ and $K'_{r\theta}$ from the frequencies of modes $\BIIu$, $\Ag^1$ and $\Ag^2$. Finally, we get 
$K_r - K_{rr}$ from mode $\BIIg$. This procedure guarantees that the VFM yields for given $\omega_{\BIIu}$,
$K_{\theta}$ and $K_{r\theta}$ the desired frequencies for the Raman active modes.

\begin{table}[b!]
    \centering
    \small
\caption{Phonon frequencies at the $\Gamma$-point in ${\rm cm^{-1}}$. Those six modes are
Raman-active. One additional silent mode
($\Au$) and the two infrared modes ($\BIu$, $\BIIu$) are not listed here. See supplement for a complete list}
\label{tab:freqs}
\begin{tabular*}{0.48\textwidth}{@{\extracolsep{\fill}}ccccccc}
    mode        & \cite{sush85}   & \cite{suue+81}  & \cite{lash79}  & \cite{akko+97} & \cite{cavi+14} & mean \\
    \hline
    $\Ag^1$	    & $365$     & $362$ & $365$ & $362$ & $362.5$ & $363.3 \pm 1.6$\\
    $\Ag^2$	    & $470$     & $467$ & $471$ & $466$ & $467.1$ & $468.2 \pm 2.2$\\
    $\BIg$	    & $197$     & $194$ & $195$ & $192$ & $-$     & $194.5 \pm 2.1$\\
    $\BIIg$	    & $442$     & $439$ & $441$ & $439$ & $439.8$ & $440.2 \pm 1.3$\\
    $\BIIIg^1$  & $223$     & $-$   & $230$ & $228$ & $-$     & $227 \pm 3.6$\\
    $\BIIIg^2$  & $440$     & $-$   & $436$ & $-$   & $-$     & $438 \pm 2.8$\\
\end{tabular*}
\end{table}
Comparing Raman frequencies from different experiments\cite{lash79,suue+81,akko+97,cavi+14} one finds that the respective values agree within $8{\rm cm^{-1}}$.
Consequently, we can use the average frequencies to calculate the VFM parameters. All Raman-active frequencies are shown in Table \ref{tab:freqs}.

As it was pointed out previously\cite{kaka+86}, the VFM given by Eq.\ \eqref{eq:vfm_en} does not involve polarization and thus
it has difficulties to correctly describe infrared modes. Therefore, we do not use $\omega_{\BIIu}\approx470{\rm cm^{-1}}$ (as reported in Refs.\  \cite{lash79,sush85}) directly.
Instead we optimize the elastic properties by tuning $\omega_{\BIIu}$,
$K_{\theta}$ and $K_{r\theta}$.

\subsection{Elastic properties}
The elastic properties determine the behavior of the acoustic phonon branches close to the $\Gamma$-point. In each direction there are three acoustic branches, two of which are linear in the wave-vector and one which is quadratic. The linear branches relate to the energetics of stretching and shearing the unit cell (i.e., to changing the size and shape). These branches are determined by the elastic constants $C_{ij}$ $(i,j=1,2,6)$. The quadratic branches relate to bending the phosphorene membrane (i.e., to a rotation of neighboring unit cells around the $x$- or $y$-axis) and are given by the bending rigidities $\kappa_x$ and $\kappa_y$.

To obtain the properties of the linear branches from the VFM, we use the approach put forward in Ref.\ \cite{mile+16}. This approach
relates the VFM to the elastic stretching energy-density in terms of the strain tensor via a minimization procedure.
Once the energy-density is known, one can calculate the elastic constants\cite{wepe14}, and from those one finds
the Young's moduli, $Y_{x/y}$, the shear modulus, $G_{xy}$, the Poisson ratios $\nu_{xy}$ and $\nu_{yx}$, and the sound
velocities $c_{xx}$, $c_{yy}$ and $c_{xy}$. Moreover, we can also calculate the Poisson ratios with respect to changes of the
thickness, $\nu_{xz}$ and $\nu_{yz}$.

\begin{figure}[t!]
  \centering
  \includegraphics[width=0.4\textwidth]
             {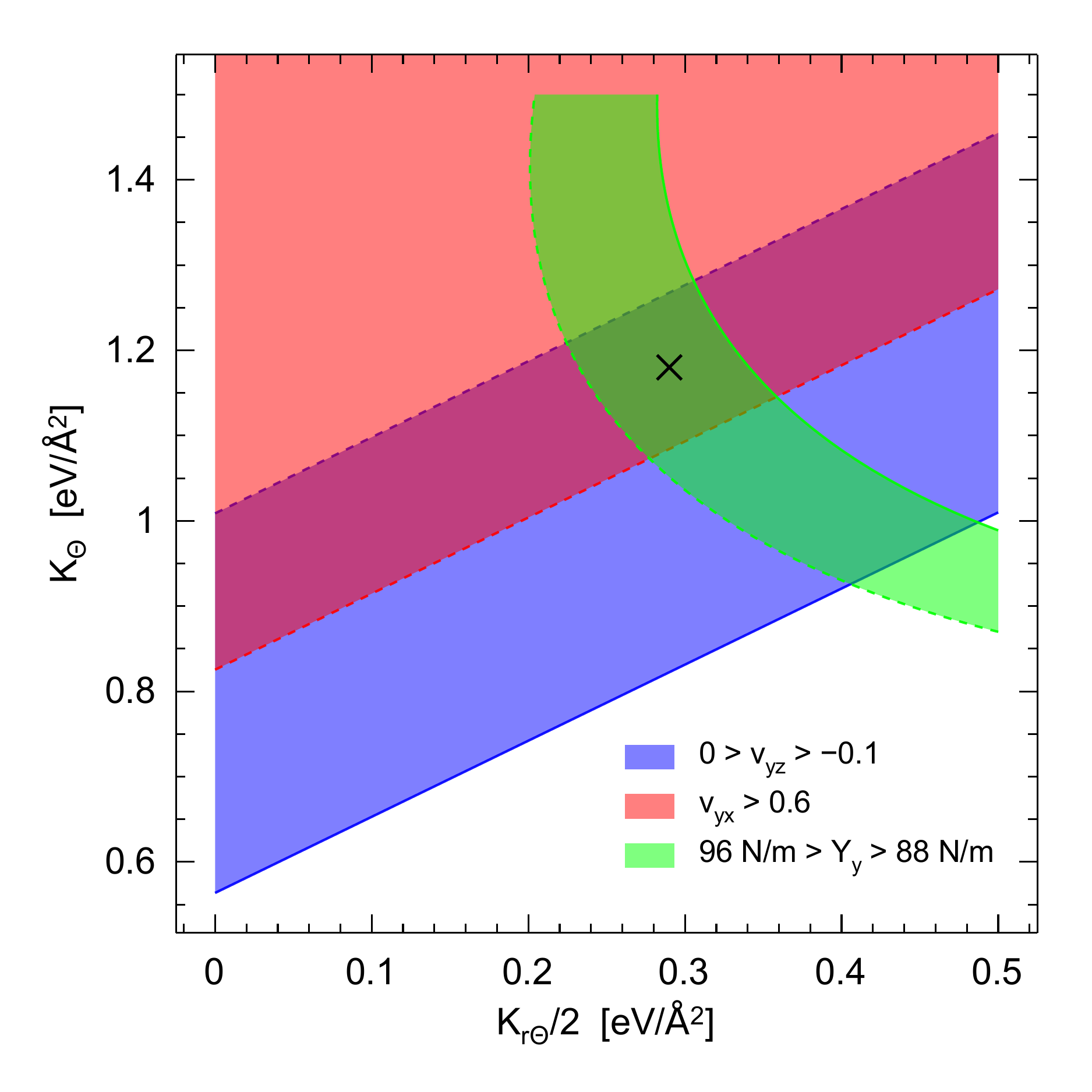}           
  \caption{Parameter space spanned $K_{\theta}$ and $K_{r\theta}$. The colored regions indicate parameters where the given constraints on $Y_y$, $\nu_{xy}$ and $\nu_{yz}$ are fulfilled. There is one region where all three regions are overlapping. The cross denotes
  the values used for the VFM.}
  \label{fig:param_est}
\end{figure}

We start by observing that the shear modulus $G_{xy}$, and thus $c_{xy}$, depends only on the frequency $\omega_{\BIIu}$.
By choosing $\omega_{\BIIu}=483{\rm cm^{-1}}$ we get $G_{xy}=22.4{\rm N/m}$ ($c_{xy}=3949{\rm m/s}$), which agrees well with recent theoretical results\cite{wepe14,sali+14,waku+15}.
Next, we impose constraints on $Y_y$, $\nu_{xy}$ and $\nu_{yz}$, which reflect recent theoretical findings\cite{wepe14,elkh+15,qiko+14}.
In Fig.\ \ref{fig:param_est} the constraints are represented by colored regions in the parameter space spanned by $K_{\theta}$ and $K_{r\theta}$. As one can see there is a region where all constraints are fulfilled. In this part we pick the values
$K_{\theta}=1.18{\rm eV}/$\AA$^2$ and $K_{r\theta}=2\times0.29{\rm eV}/$\AA$^2$, which are indicated by the black cross in Fig.\ \ref{fig:param_est}.
Now we have determined all nine VFM parameters and their values are given in Table \ref{tab:vfm} along with the original parameters\cite{kaka+82}. The new values for bond stretching and angle bending are comparable to the old ones, the largest difference is found for
the bond-bond and bond-angle correlation parameters, $K_{rr'}$ and $K'_{r\theta}$. 
The resulting elastic properties are displayed in Table \ref{tab:elastic} and compared to values found from \emph{ab initio} 
calculations. The values for the Young's modulus $Y_y$, the shear modulus $G_{xy}$ and the Poisson ratios agree well
with the DFT results. In particular, we find a negative out-of-plane Poisson ratio $\nu_{yz}$ as
reported before\cite{jipa14,elkh+15}. However, the Young's modulus $Y_x$ is about $30\%$ smaller than typically reported values,
but similar to the value found for the original VFM.

\begin{table*}[bt!]
    \centering
    \small
\caption{VFM parameters in ${\rm eV}/$\AA$^2$}
\label{tab:vfm}
\begin{tabular*}{0.9\textwidth}{@{\extracolsep{\fill}}lccccccccc}
        & $K_r$ & $K'_r$ & $K_{\theta}$ & $K'_{\theta}$ & 
          $K_{rr'}$ & $K'_{rr'}$ & $K_{r\theta}$ & $K'_{r\theta}$ & $K''_{r\theta}$ \\
    \hline
    Ref.\ \cite{kaka+82}	& $9.97$  &  $9.46$   & $1.08$ & $0.93$     & $1.11$      & $1.11$    & $0.72$ & $0.72$    & $0.72$ \\
    this work	            & $11.17$ & $10.3064$ & $1.18$ & $0.9259$ & $-0.6763$ & $1.2449$ & $0.58$ & $1.932$ & $0.797$ \\
    \hline
\end{tabular*}
\end{table*}
\begin{table}[bt!]
    \centering
    \small
\caption{Young's moduli, shear modulus and Poisson ratios. The former are given
in units of ${\rm N/m}$ and the latter are dimensionless}
\label{tab:elastic}
    \begin{tabular*}{0.48\textwidth}{@{\extracolsep{\fill}}lccccccc}
                            & $Y_x$     & $Y_y$     & $G_{xy}$  & $\nu_{yx}$ & $\nu_{xy}$   & $\nu_{xz}$ & $\nu_{yz}$\\
\hline
VFM  \cite{kaka+82}         & $17.4$    & $93.6$    & $20.8$    & $0.27$     & $0.051$      & $0.35$ & $0.10$ \\
\hline
this work                   & $16.2$    & $90.3$    & $22.4$    & $0.62$     & $0.11$       & $0.37$ & $-0.08$ \\
\hline
DFT \cite{elkh+15}          & $26$      & $88$      & -         & $0.81$ & $0.24$           & $0.21$ & $-0.09$ \\
DFT \cite{wepe14}           & $24$      & $92$      & $22.8$    & $0.62$ & $0.17$           & -  & -  \\
DFT \cite{waku+15}          & $23$      & $92.3$    & $22.4$    & $0.703$ & $0.175$         & -  & - \\
    \end{tabular*}
\end{table}

The bending rigidities $\kappa_x$ and $\kappa_y$ are obtained from the VFM by extending the approach of Ref.\ \cite{mile+16} to a curved phosphorene configuration. We induce a curvature of the phosphorene in the $x-$ or $y-$ direction by constraining bond vectors separated by a lattice vector along the principal direction of curvature to be rotated with respect to each other by an angle $\phi=a_i/\xi$ $(i=1,2)$ where $\xi$ is the radius of curvature. We then minimize the VFM energy under those constraints and obtain the energy required to induce a curvature on the phosphorene. Using this procedure, we obtain the bending rigidities $\kappa_x=1.1\,{\rm eV}$ and $\kappa_y= 7.4\,{\rm eV}$ for our set of VFM parameters.

\subsection{Phonon dispersion}
\begin{figure}[t!]
  \centering
  \includegraphics[width=0.23\textwidth]
             {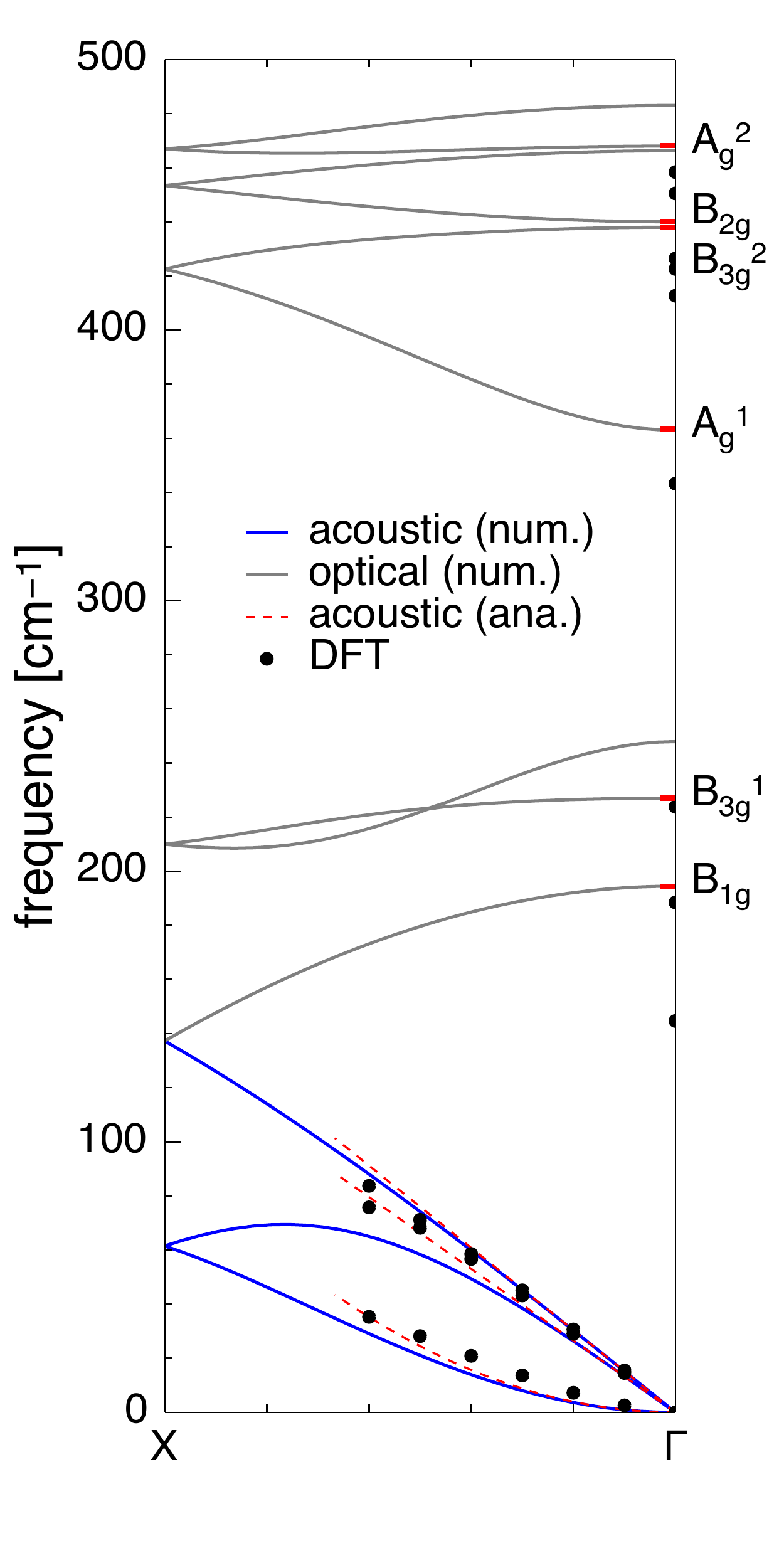}           
  \includegraphics[width=0.23\textwidth]
             {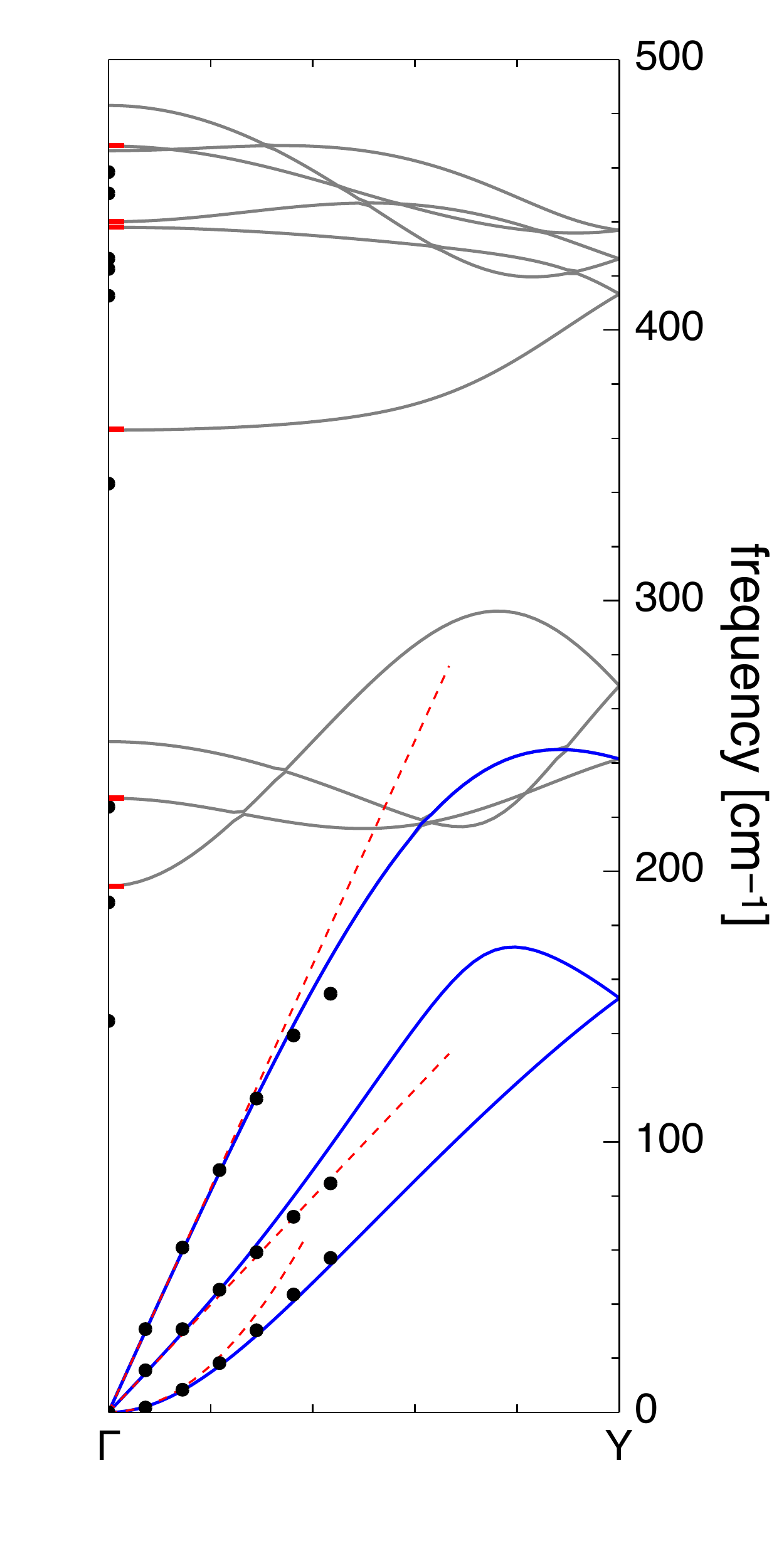}           
             \caption{Phonon dispersion of phosphorene numerically calculated from the VFM is indicated by solid lines. The dashed lines show the behavior of the acoustic branches according to the continuum model. Filled circles are {\it ab initio} data from Ref.\ \cite{feya14a}.}
  \label{fig:dispersion}
\end{figure}
In Fig.\ \ref{fig:dispersion} we show the complete phonon dispersion calculated from the VFM. By construction the phonon frequencies at the $\Gamma$-point agree with the average frequencies given in Table \ref{tab:freqs}. The slope of the linear branches
yields the longitudinal and transversal sound velocities in the respective directions. As one can see, those slopes
coincide with the sound velocities obtained by the procedure of Ref.\ \cite{mile+16} (dashed lines). Finally, we can compare
the curvature of the quadratic branches with bending rigidities obtained above. Also here we find a very good agreement.

Additionally, Fig.\ \ref{fig:dispersion} contains data from an \emph{ab initio} calculation\cite{feya14a}. The respective phonon frequencies at the $\Gamma$-point are consistently lower than the experimental values and thus lower than the frequencies found from our VFM. This is consistent with the fact that typically DFT calculations yield larger unit cells, which are thus stretched compared to our cell. Applying uniaxial strain leads to lower phonon frequencies\cite{feya14a}.
For clarity, we only show the dispersion for small momenta. The acoustic branches agree very well with those found from the VFM. 

\subsection{Continuum model}
Denoting a local deformation of the sheet by the displacement vector $\vec{u}=(u(x,y),v(x,y),w(x,y))$ where $u$, $v$ and $w$ are the displacements of the local sheet coordinates in $x-$, $y-$ and $z-$direction, the we can write down the elastic energy-density valid for long wavelengths
\begin{equation}
\label{eq:energydensity}
\mathcal{E}=\frac{1}{2} \left(\sqrt{\kappa_x} \partial^2_x w + \sqrt{\kappa_y} \partial^2_y w\right)^2 + \frac{1}{2} \sigma_{ij} \epsilon_{ij} \;,
\end{equation}
where the stress tensor $\sigma_{ij}$ is related to the strain tensor $\epsilon_{ij}$ via \cite{lali86}
\begin{align}
\label{eq:Hooke}
\sigma_{xx} = &(1-\tilde{\nu}^2)^{-1}\left( Y_x \epsilon_{xx} + \tilde{\nu} \tilde{Y} \epsilon_{yy}\right)\;, \nonumber \\
\sigma_{yy} = &(1-\tilde{\nu}^2)^{-1}\left( Y_y \epsilon_{yy} + \tilde{\nu} \tilde{Y} \epsilon_{xx}\right)\;, \nonumber \\
\sigma_{xy} = & 2 G_{xy} \epsilon_{xy}\;,
\end{align} 
with $\tilde{\nu} \equiv \sqrt{\nu_{xy} \nu_{yx}}$ and $\tilde{Y} \equiv \sqrt{Y_{x} Y_{y}}$.
Further, the strains are connected to the displacements by \cite{lali86}
\begin{align}
\label{eq:strains}
\epsilon_{xx} = & \partial_x u + (1/2) \left( \partial_x w\right)^2\;, \nonumber \\
\epsilon_{yy} = & \partial_y v + (1/2) \left( \partial_y w\right)^2\;, \nonumber \\
\epsilon_{xy} = & (1/2) \left( \partial_y u + \partial_x v + \partial_x w \partial_y w\right)\;.
\end{align}
From these equations one obtains the compatibility equation
\begin{equation}
\label{eq:comp}
\partial^2_y \epsilon_{xx} + \partial^2_x \epsilon_{yy} - 2\partial_x \partial_y \epsilon_{xy}= (\partial_x \partial_y w)^2-(\partial^2_x w) (\partial^2_y w)\;.
\end{equation}
This equation must be fulfilled for any physical strain configuration in order to ensure continuity in the displacement fields.

Due to the anisotropy of phosphorene, the energetics of stretching and bending has a directional dependence. 
For a sheet strained at an angle $\theta$ with respect to the symmetry axes of the unstrained phosphorene sheet, the directionally dependent Poisson ratios $\nu(\theta)$ and tensile strength $T(\theta)$ are described in Ref.\ \cite{waku+15}. Additionally, we find a directionally dependent bending rigidity
\begin{equation}
\kappa(\theta) = (\sqrt{\kappa_x} \cos^2(\theta) + \sqrt{\kappa_y}\sin^2(\theta))^2 \;.
\end{equation}

The equations of motion for the local sheet deformations are obtained from the Lagrangian $\mathcal{L} = \frac{\rho}{2} \dot{\vec{u}}\cdot\dot{\vec{u}} - \mathcal{E}$, where $\rho$ is the two-dimensional mass density of phosphorene
and $\dot{\vec{u}}$ denotes the time-derivative of the displacement vector.
This leads to the F\"{o}ppl-von Karman equations \cite{lali86}
\begin{align}
\rho \ddot{u} ={}& \partial_x \sigma_{xx} + \partial_{y} \sigma_{xy}\;, \nonumber \\
\rho \ddot{v} ={}& \partial_y \sigma_{yy} + \partial_{x} \sigma_{xy}\;, \nonumber \\
\rho \ddot{w} ={}& -\left(\kappa_x \partial^4_x w + \kappa_y \partial^4_y w + 2 \sqrt{\kappa_x \kappa_y} \partial^2_x \partial^2_y w\right) + \nonumber \\
&\partial_x\left( \sigma_{xx} \partial_x w + \sigma_{xy} \partial_y w\right) +\partial_{y}\left(\sigma_{yy}\partial_y w +\sigma_{xy}\partial_x w\right) \nonumber \\
    &+ P_z\;, 
\label{eq:eom}
\end{align}
where $P_z$ is an externally applied pressure in $z$-direction. These equations need to be augmented with the proper boundary conditions. For a free boundary with normal unit vector $\hat{n}$, one requires $\underline{\sigma}\cdot \hat{n} = 0$ and $\partial^2 w/\partial\hat{n}^2 =0$. For a fixed boundary $\Omega$ one instead requires $u(\Omega) = v(\Omega) = w(\Omega) =0$, and $\partial w/\partial \hat{n} =0$. 

The equations of motion can be simplified when out-of-plane vibrations or deformations are considered, since the relaxation times for the in-plane motion typically is at least an order of magnitude faster than the relaxation time for the out-of-plane vibrations. Then, the stresses are assumed to always be in equilibrium, so that the time-derivatives in the first two equations can be dropped. Further, we introduce the Airy stress function $\chi$ as usual\cite{lali86}
\begin{equation}
\label{eq:airy}
\sigma_{xx} = \partial^2_y \chi,\;\;\;\sigma_{yy} = \partial^2_x \chi, \;\;\; \sigma_{xy} = -\partial_x\partial_y \chi\;.
\end{equation}
It is easily verified that this Ansatz for the stresses satisfies the equilibrium stress equations for arbitrary $\chi$. Requiring in addition that the compatibility equation \eqref{eq:comp} is satisfied one finds for the stress function\cite{le68}
\begin{align}
\label{eq:airyeq}
(1/Y_y)& \partial^4_x \chi + (1/Y_x) \partial^4_y \chi 
+ \left((1/G_{xy}) -2(\tilde{\nu}/\tilde{Y}) \right) \partial^2_x\partial^2_y \chi=\nonumber \\
&(\partial_x \partial_y w)^2-(\partial^2_x w) (\partial^2_y w)\;.
\end{align}
Moreover, the equation of motion for the out-of-plane displacement becomes
\begin{align}
\label{eq:w}
\rho \ddot{w} = & -\left(\kappa_x \partial^4_x w + \kappa_y \partial^4_y w + 2 \sqrt{\kappa_x \kappa_y} \partial^2_x \partial^2_y w\right) + \nonumber \\
&(\partial^2_y \chi) (\partial^2_x w) + (\partial^2_x \chi) (\partial^2_y w) -2 (\partial_x\partial_y \chi) (\partial_x\partial_y w) + P_z\;.
\end{align}
Together, equations \eqref{eq:airyeq} and \eqref{eq:w} constitute the relevant equations of motion of a suspended phosphorene sheet.

\subsection{Suspended phosphorene drum}
\begin{figure}[b!]
	\centering
        \includegraphics[width=0.4\textwidth]
	        {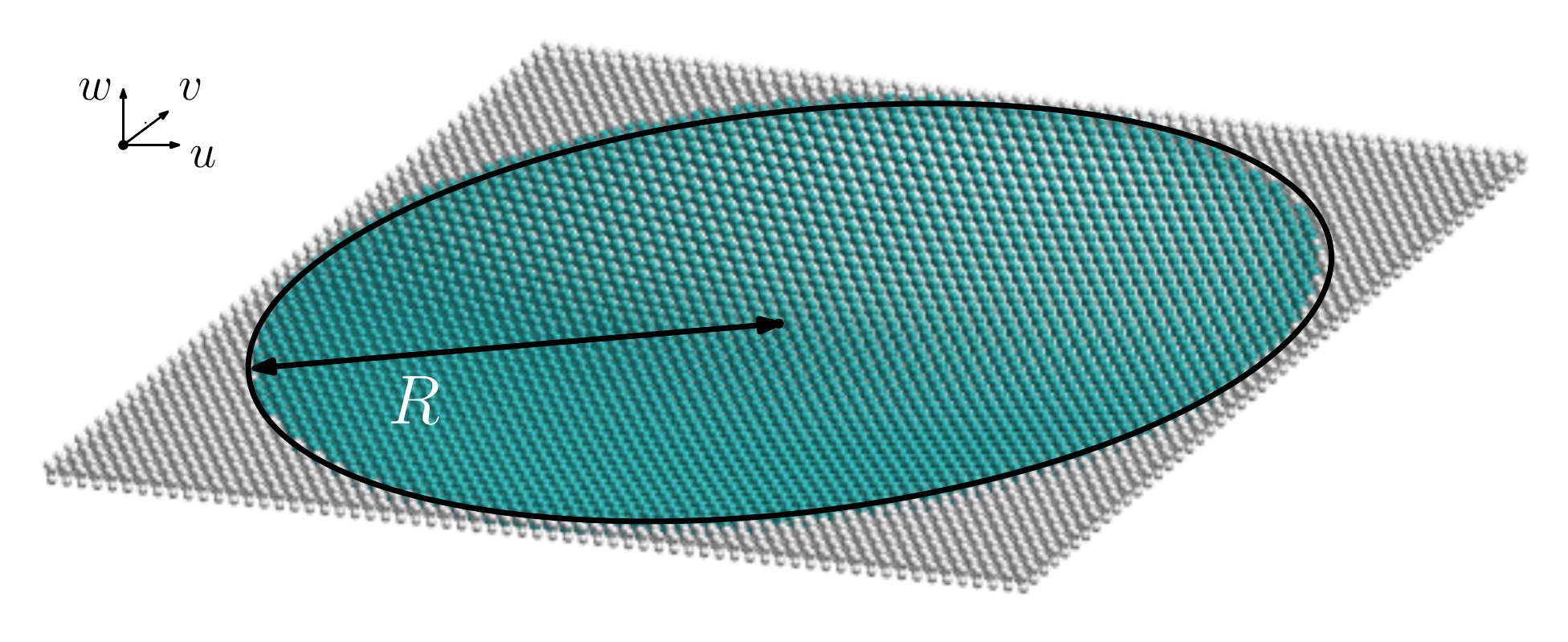}
    \caption{Sketch of a suspended phosphorene drum.}
	\label{fig:drum_sketch}
\end{figure}
The effective model for long wavelength deformations given by Eqs.\ \eqref{eq:airyeq} and \eqref{eq:w}, allows us to describe properties of macroscopic phosphorene configurations not directly accessible within a purely microscopic approach. As a specific example we consider the out-of-plane displacements of a suspended phosphorene sheet with a circular region of suspension\cite{waji+15} of radius $R$ as shown in figure \ref{fig:drum_sketch}. The sheet is subject to a static, uniform vertical pressure $P_z$, and the resulting deformations are found by solving \eqref{eq:airyeq} and \eqref{eq:w} and setting the time-derivatives to zero. In order to more closely mimic an experimental realization, the sheet is additionally subject to a uniform pre-strain $\epsilon_{xx}^0=\epsilon_{yy}^0 = \epsilon$ and no shear $\epsilon_{xy}^0=0$.

Before attempting at a solution, we note that the first three terms in \eqref{eq:w} are related to bending of the membrane, and the other three terms are related to stretching of the membrane. The relative importance of these two terms is determined by the size of the drum and of the pre-strain. To see this, we introduce dimensionless coordinates $\tilde{x} = x/R$, $\tilde{y}=y/R$ and displacements $\tilde{w}=w/w_0$, $\tilde{u} = (uR/w_0^2)$, $\tilde{v} = (vR/w_0^2)$, where $w_0$ is the deflection at the center of the drum. The deformation energy of such a drum becomes to lowest order in $\tilde{w}$ 
\begin{equation}
E=(w_0^2/2) \left[\epsilon T_y I_1 + \kappa_y/R^2 I_2\right]\;,
\end{equation}
where $I_1 \equiv \int d\tilde{x} d\tilde{y} \left((\partial_{\tilde{y}} \tilde{w} )^2 + (T_x/T_y) (\partial_{\tilde{x}} \tilde{w} )^2\right)$ and $I_2\equiv \int d\tilde{x} d\tilde{y} \left(\partial^2_{\tilde{y}} \tilde{w} + \sqrt{\kappa_x/\kappa_y} \partial^2_{\tilde{x}} \tilde{w} \right)^2$. 
The integrations are taken over the unit disc and $T_x =(Y_x+\tilde{\nu} \tilde{Y})(1-\tilde{\nu}^2)^{-1}$ and $T_y =(Y_x+\tilde{\nu} \tilde{Y})(1-\tilde{\nu}^2)^{-1}$. For $\sqrt{\epsilon} R < \sqrt{(\kappa_y I_2)/(T_y I_1)}$ the energy is dominated by the bending energy, otherwise by the stretching contribution. Using the values for the elastic parameters extracted in the previous section, we find that the length scale defined on the right-hand side of the above inequality is $\sqrt{(\kappa_y I_2)/(T_y I_1)} \approx 3.8$ \AA{}. In other words, for drums larger than $3.8\text{\AA}/\sqrt{\epsilon}$ the stretching energy dominates, for drums smaller than this length scale the bending energy dominates for small deflections. 

In the bending regime, the out-of-plane deformations decouple from the in-plane deformations. The equation for the deflection becomes
\begin{equation}
\frac{P_z R^4}{w_0} = \left({\kappa_x} \partial^4_{\tilde{x}} + \kappa_y\partial^4_{\tilde{y}} + 2 \sqrt{{\kappa_x}{\kappa_y}} \partial^2_{\tilde{x}} \partial^2_{\tilde{y}} \right)\tilde{w}\;,
\end{equation}
with boundary conditions $\tilde{w}|_{\tilde{x}^2+\tilde{y}^2=1}=0$ and $\partial \tilde{w}/\partial\hat{n} |_{\tilde{x}^2+\tilde{y}^2=1}=0$. It is easily verified that the Ansatz $\tilde{w} = (1-\tilde{x}^2-\tilde{y}^2)^2$ satisfies this equation with the proper boundary conditions. By insertion one obtains
\begin{equation}
\label{eq:Bending}
    \frac{w_0}{R} = \frac{R^3 P_z}{64 \kappa_{\rm eff} } \; \; (\rm{bending})\;,
\end{equation}
where $\kappa_{\rm eff} = \left(3\kappa_x +3\kappa_y + 2\sqrt{\kappa_x \kappa_y}\right)/8 \approx 3.9\,{\rm eV}$ for the bending
rigidities found from the VFM. 

In the opposite limit, the situation is more complicated. For drums with radius $R>\sqrt{(\kappa_y I_2)/(T_y I_1)}\epsilon^{-1/2}$ the elastic energy is dominated by the stretching contribution. However, close to the boundary, there is a boundary layer of thickness $r = \sqrt{(\kappa_y I_2)/(T_y I_1)}$ where the bending energy still dominates. Due to this boundary layer, ignoring the bending contribution in the equations of motion gives an accurate description of the drum shape only far away from the boundary. With this in mind, we require the Ansatz for the deformation in the stretching regime to obey $\tilde{w}|_{\tilde{x}^2+\tilde{y}^2=1}=0$, without restrictions on the derivative at the boundary. In the linear regime of small deformations, the induced stress is negligible compared to the stress due to the pre-strain $\epsilon$, in which case the out-of-plane deformation also decouples from the in-plane deformations. The equation of equilibrium is
\begin{equation}
    \frac{P_z R^2}{w_0} = \left({\sigma_{xx}^0} \partial^2_{\tilde{x}} + \sigma_{yy}^0\partial^2_{\tilde{y}} \right)\tilde{w}\;,
\end{equation}
with $\sigma_{yy}^0 = \epsilon(Y_y+\tilde{\nu} \tilde{Y})(1-\tilde{\nu}^2)^{-1}$ and $\sigma_{xx}^0 = \epsilon (Y_x+\tilde{\nu} \tilde{Y})(1-\tilde{\nu}^2)^{-1}$. Now, the Ansatz $\tilde{w} = 1-(\tilde{x}^2 + \tilde{y}^2)$ satisfies this equation, yielding
\begin{equation}
    \frac{w_0}{R} = \frac{(1-\tilde{\nu}^2)}{2\epsilon \left(Y_x + Y_y + 2\tilde{\nu} \tilde{Y}\right)} R P_z = \frac{R P_z}{4 \epsilon T_{\rm eff}}  \quad (\rm{stretching,\;linear})\;,
\end{equation}
with $T_{\rm eff}\equiv (1-\tilde{\nu}^2)^{-1} \left(Y_x + Y_y + 2\tilde{\nu} \tilde{Y}\right)/2$. For $Y_x = Y_y =\tilde{Y}\equiv Y$ one obtains $T_{\rm eff} = Y/(1-\nu)$ and one recovers the expression for isotropic plates\cite{ermi+13}.

Regardless of whether the drum is dominated by bending energy or stretching energy at small deformations, at sufficiently large deformations the stress induced by the deformation needs to be taken into account, rendering the problem nonlinear. To model this regime, we again use the Ansatz $\tilde{w} = 1-(\tilde{x}^2 + \tilde{y}^2)$ and insert it into the Airy stress equation \eqref{eq:airyeq}. The solution is then given by $\chi = A x^4+B y^4 + C x^2 y^2 + D x^2 + E y^2 + \chi_0$, where $\chi_0$ is an arbitrary polynomial in $x$ and $y$ of order less than two. The coefficients of the Airy function are obtained by insertion into Eq.\ \eqref{eq:airyeq} and from the boundary conditions for the in-plane displacements. From \eqref{eq:airy} we find the corresponding stresses and insert them into the equation of equilibrium of the out-of-plane deformation to obtain
\begin{equation}
\label{eq:NLStretch}
\left(\frac{w_0}{R}\right)^3 = \frac{(1-\tilde{\nu}^2)}{2 T_{\rm eff}} R P_z \quad (\rm{stretching,\;nonlinear})\;.
\end{equation}
Using the values of the elastic properties given in Table \ref{tab:elastic} we find $T_{\rm eff}\approx 67.9\;{\rm N/m}$ and $\tilde{\nu} \approx 0.26$.

From this result we find that the crossovers from the linear stretching/bending to nonlinear stretching regime occurs at the crossover deflection $w_0^c$,
\begin{align}
    w_0^c =& \sqrt{2} \left( \frac{(1-\tilde{\nu}^2)}{T_{\rm eff}} \kappa_{\rm eff}\right)^{1/2} \approx 5.2\, \text{\AA}\quad (\rm{bending})\;,\nonumber \\
    w_0^c = & \sqrt{\epsilon}R \sqrt{2 (1-\tilde{\nu}^2)} \approx 1.4 R\sqrt{\epsilon}\quad (\rm{stretching})\;.
\end{align}
\begin{figure*}[t!]
	\centering
    \begin{minipage}[b]{0.6\textwidth}
    	\centering
        \includegraphics[width=\textwidth]
        	{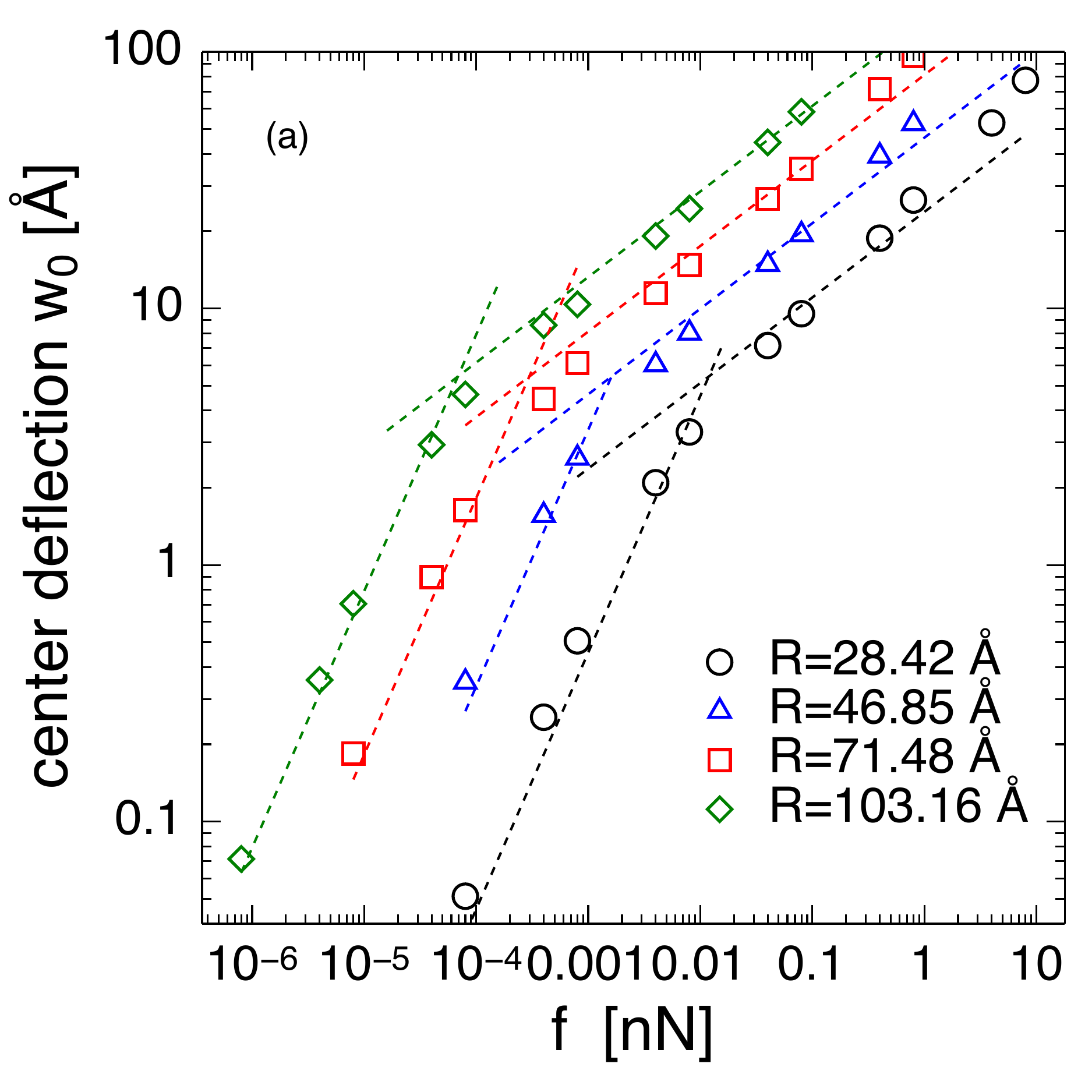}
    \end{minipage}%
    \begin{minipage}[b]{0.3\textwidth}
        \centering
	    \includegraphics[width=\textwidth]
	        {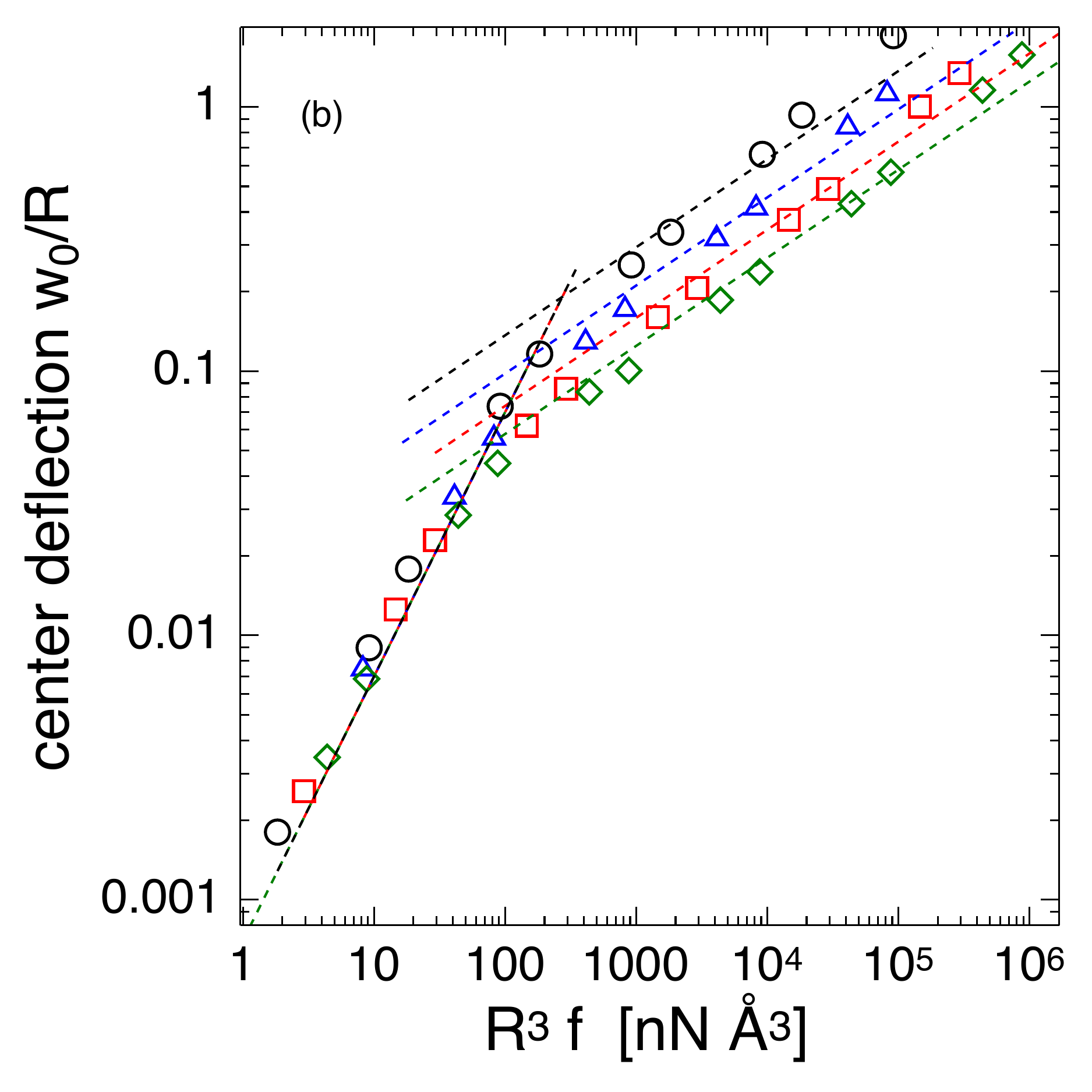}\\
        \includegraphics[width=\textwidth]
	        {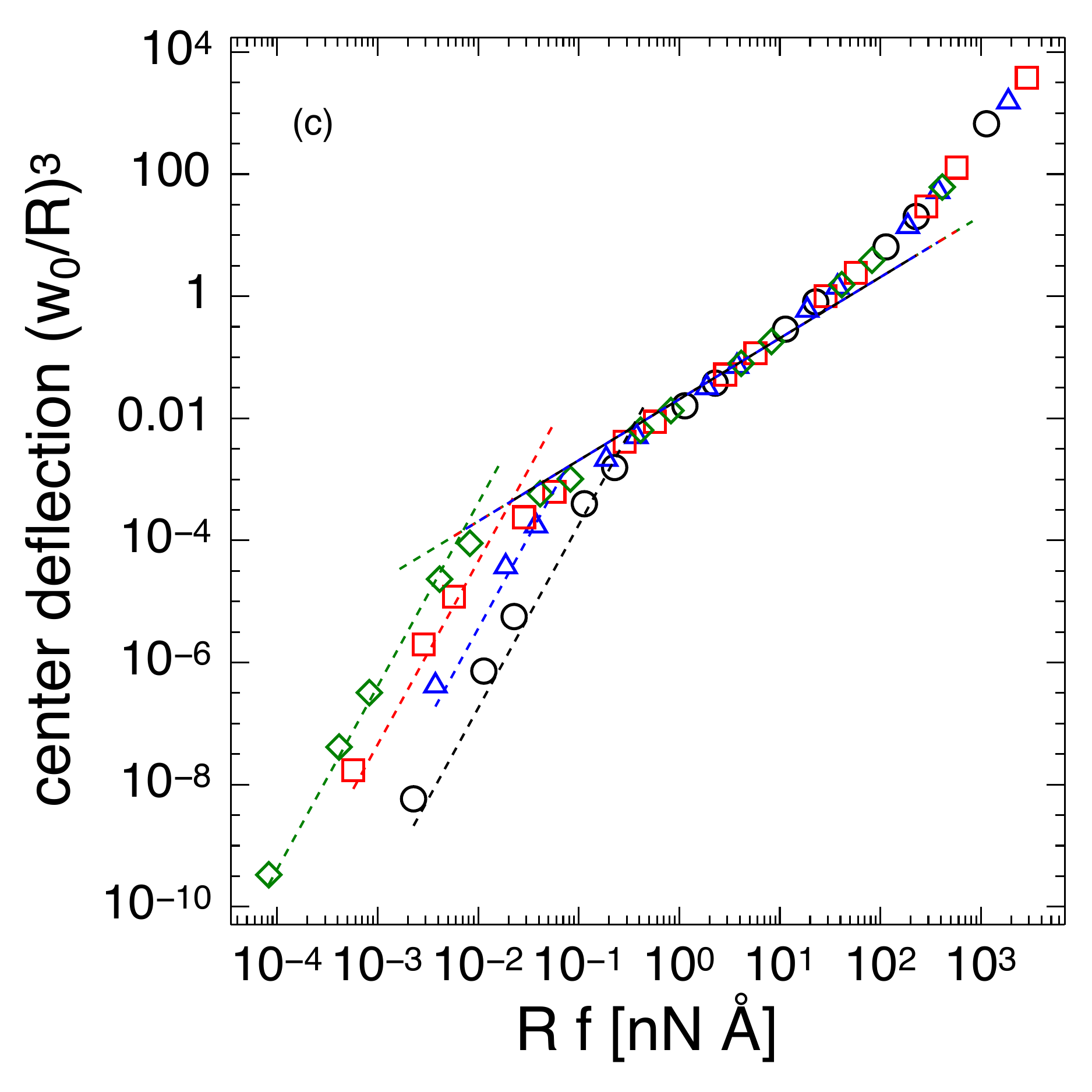}
    \end{minipage}
	\caption{(a-c) Deflection $w_0$ at the center of the drum as a function of the
   applied force per atom $f$ for different radii. Symbols denote numerical results, while dashed lines indicate the
behavior according to Eqs.\ \eqref{eq:Bending} and \eqref{eq:NLStretch}. (b) Bending regime for small forces. (c) Nonlinear stretching regime for large forces. After scaling the axes the curves for different radii collapse in the respective regimes.}
	\label{fig:drum}
\end{figure*}
To verify our results we employ the VFM model and perform an atomistic simulation of the pressurized drum with radii between $28.42$ \AA{} and $103.16$ \AA{}. In Fig.\ \ref{fig:drum}(a) the resulting deflection at the center of the drum is shown as a function of
the applied force per atom. This force $f$ corresponds to a pressure $P_z=4 f/A_{\rm uc}$, where $A_{\rm uc}$ is the area of the unit cell. Since no pre-strain is applied to the drum, its shape is dominated by the bending energy for sufficiently low pressure. We confirm this by plotting $w_0/R$ against $R^3 P_z$ and find that the numerical results fall onto a single line for small forces in accordance with Eq.\ \eqref{eq:Bending} (figure \ref{fig:drum}(b)). For larger deformations there is a crossover to the nonlinear stretching regime. The results now collapse onto a single line after rescaling the pressure by the radius, in accordance with Eq.\ \eqref{eq:NLStretch} (figure \ref{fig:drum}(c)).

The strain induced by the deformation is of interest for strain-engineering applications and is readily calculated from the Airy stress function using Eqs. \eqref{eq:airy} and \eqref{eq:Hooke}. We find that the induced strain in radial coordinates ($\tilde{x}=r \cos(\theta)$, $\tilde{y}=r \sin(\theta)$, $0\leq r\leq 1$ and $0\leq\theta<2 \pi$) is given by
\begin{align}
    \epsilon_{rr}(r,\theta) =& \frac{w_0^2}{R^2}\left(4 r^2 + (1-3 r^2) \left(a+b \cos(2 \theta)\right)\right) \;, \nonumber \\
    \epsilon_{\theta\theta}(r,\theta) =& \frac{w_0^2}{R^2}(1-r^2)(a - b \cos(2\theta))\;, \nonumber \\
    \epsilon_{r\theta}(r,\theta) =& \frac{w_0^2}{R^2} (2 r^2-1) b \sin(2\theta) \;,
\end{align}
where the constants $a$ and $b$ are functions of the elastic constants as given in the supplement. The appearance of the parameter $b$ breaks the radial symmetry of the induced strains. As we show in the supplement, this parameter is proportional to the difference between the Young's moduli in the two principal directions, $Y_x-Y_y$. Thus, the material anisotropy gives rise to a non radially symmetric strain distribution when a phosphorene sheet is deformed in a radially symmetric fashion.

As a final application of the model, we estimate the resonance frequency of the fundamental oscillation mode of the phosphorene drum in the bending and stretching regime\cite{wafe15}. Using the same Ansatz for the mode shape as for the static deflection one can estimate the mode frequencies by projecting the mode shape onto Eq.\ \eqref{eq:w}. By this procedure one finds the frequencies
\begin{align}
\omega^2_{\rm bend}& \approx \frac{320 \kappa_{\rm eff} }{3\rho R^4} \quad ({\rm bending})\;,\nonumber \\
    \omega^2_{\rm stretch}& \approx \frac{6 T_{\rm eff}}{\rho R^2}\left(\epsilon + \frac{(1-\tilde{\nu}^2)}{2}\left(\frac{w_0}{R}\right)^2\right) \quad ({\rm stretching})\;. 
\end{align}
In deriving the mode frequency in the stretching regime the stress induced by the deformation was replaced by its spatial average. Note that these estimates are upper bounds to the true fundamental mode frequencies, since the mode shape Ansatz is only approximate.

\section{Conclusions}
In conclusion, we have developed a valence force model for phosphorene. The resulting phonon dispersion and the elastic
constants are in good agreement with experimentally obtained values and \emph{ab initio} results. We used the elastic
constants and the bending rigidities to provide a complete continuum mechanics model, which facilitates the
description of nanomechanical\cite{micr+14,waji+15,wafe15} and phononic\cite{miis+14} applications. In this context, we have studied a pressurized, suspended phosphorene
drum and give analytical expressions for its deflection and induced strain as a function of the applied pressure. Such configurations
are of great interest since they provide means to introduce a controllable strain into the sheet. For example, it has recently been proposed
that radially symmetric deformations of phosphorene may be exploited through a strong anisotropic inverse funnel effect\cite{sapa+16}. Interestingly, we found that while the deformation is radially symmetric, due to the anisotropy of phosphorene the induced strains do not obey
radial symmetry. This implies that in order to correctly model strains in deformed phosphorene sheets, one must use the anisotropic Airy equation \eqref{eq:airyeq} for the deformation explicitly, as the induced strain does in general not obey the same symmetries as the deformation.
Our continuum results also apply to other anisotropic single-layer materials, which renders them a good starting point for
investigating a variety of topical questions.

\section*{Acknowledgements}
We gratefully acknowledge R.\ Fei and L.\ Yang for providing data for Fig.\ \ref{fig:dispersion}.



\end{document}